\documentclass[lettersize,journal]{IEEEtran}
\usepackage{amsmath,amsfonts}
\usepackage{amssymb}
\usepackage{algorithm}
\usepackage{algorithmicx}
\usepackage[noend]{algpseudocode}
\usepackage{xcolor}
\usepackage{array}
\usepackage{textcomp}
\usepackage{stfloats}
\usepackage{color}
\usepackage{url}
\usepackage{verbatim}
\usepackage{graphicx}
\usepackage{cite}
\usepackage{amssymb}
\usepackage{kotex}
\usepackage{booktabs}
\usepackage{tabularx}
\usepackage{threeparttable}
\usepackage{multirow}
\usepackage{changepage}
\usepackage{subcaption}
\usepackage{enumitem}
\usepackage{comment}
\usepackage{pifont}

\begin{document} 
\title{NWP-based Atmospheric Refractivity Modeling and Fast \& Stable Non-uniform Plane Wave Ray-Tracing Simulations for LEO Link Analysis}
\author{Bowoo Jang,~\IEEEmembership{Student Member,~IEEE}, Jun Heo,~\IEEEmembership{Member,~IEEE}, Yong Bae Park,~\IEEEmembership{Senior Member,~IEEE}, and Dong-Yeop Na,~\IEEEmembership{Member,~IEEE,}
\thanks{B. Jang, J. Heo, and D.-Y. Na are with the Department of Electrical Engineering, Pohang University of Science and Technology (POSTECH), Pohang, Gyeongsangbuk-do, 37673, South Korea (e-mail: dyna22@postech.ac.kr).
Y. B. Park with the Department of Electrical and Computer Engineering, Ajou University, Suwon, Gyeonggi-do, 16499, South Korea.
}
\thanks{This work was supported by the Institute of Information \& Communications Technology Planning \& Evaluation (IITP) grant funded by the Korean government (MSIT) (RS-2024-00396992). {(\it Corresponding author: Dong-Yeop Na)}}}

\markboth{Journal of \LaTeX\ Class Files,~Vol.~14, No.~8, August~2021}%
{Shell \MakeLowercase{\textit{et al.}}: A Sample Article Using IEEEtran.cls for IEEE Journals}

\IEEEpubid{0000--0000/00\$00.00~\copyright~2021 IEEE}
\maketitle

\begin{abstract}
Existing low-Earth-orbit (LEO) communication link analyses face two main challenges:
(1) limited accuracy of 3D atmospheric refractivity reconstructed from sparsely sampled radiosonde data, and (2) numerical instability in previous non-uniform plane-wave ray-tracing algorithms~\cite{chang2005ray} (i.e., underflow under standard double precision), where non-uniform plane waves inevitably arise at complex-valued dielectric interfaces, is caused by extremely small atmospheric loss terms.
To address these issues, we reconstruct a high-resolution 3D complex-valued refractivity model using numerical weather prediction data, and develop a fast and numerically stable non-uniform plane-wave ray tracer. 
The method remains stable in double precision and delivers a 24-fold speedup over high-precision benchmarks.
Comparisons show that boresight-error deviations and path-loss differences between the rigorous method and the uniform-plane-wave approximation remain negligible, even under heavy precipitation. Although rays in a lossy atmosphere experience different phase- and attenuation-direction vectors---forming non-uniform plane waves---the resulting effective attenuation along the path is nearly identical to that predicted by the uniform-plane-wave model.
These findings justify the continued use of uniform-plane-wave ray tracing in practical LEO link analyses.
\end{abstract}


\section{Introduction}
\IEEEPARstart{L}{ow}-Earth-orbit (LEO) satellite communications have emerged as a key technology for non-terrestrial networks~\cite{kodheli2021IEEEcommunicationsSurveys&Tutorial, lalbakhsh2022IEEEAccess, juan2022IEEEAccess}. 
Because LEO satellites move at high velocities, seamless connectivity requires highly accurate handover protocols, which in turn demand precise modeling of atmospheric refraction---particularly severe at large incidence angles---and the resulting accumulated path loss. Consequently, rigorous electromagnetic (EM) propagation modeling under realistic atmospheric refractivity conditions is essential~\cite{pham2022determination,chakravarty2025numerical}.
\begin{figure}[t]
    \centering
    \begin{subfigure}[b]{0.48\linewidth}
        \includegraphics[width=\linewidth]{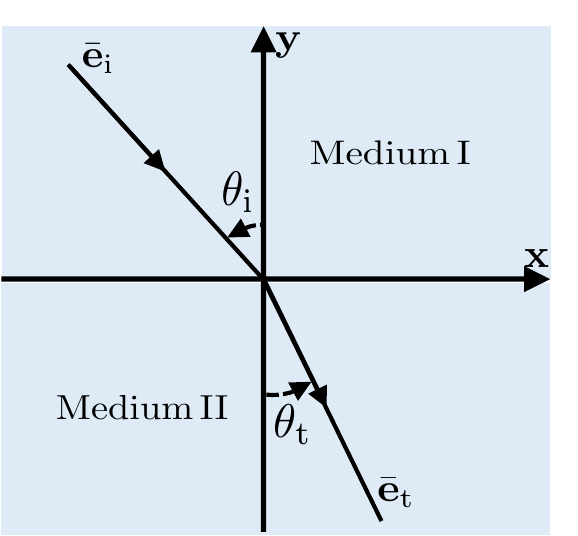}
        \caption{Uniform plane wave}
        \label{fig:HPW_incidence}
    \end{subfigure}
    \hfill
    \begin{subfigure}[b]{0.48\linewidth}
        \includegraphics[width=\linewidth]{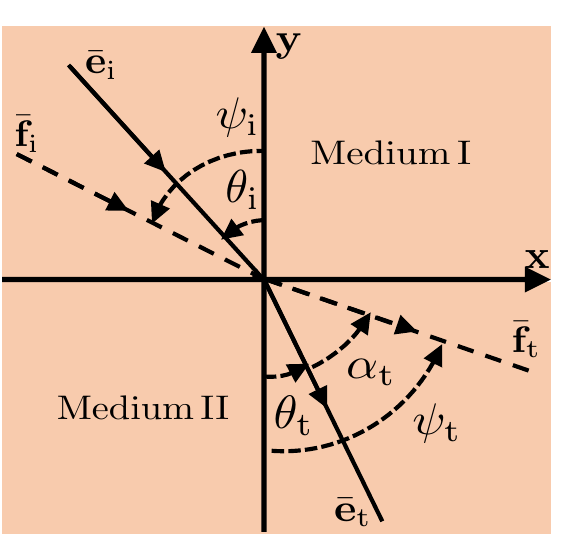}
        \caption{Non-uniform plane wave}
        \label{fig:IHPW incidence}
    \end{subfigure}
    \caption{Refraction of (a) uniform and (b) non-uniform plane waves at a dielectric interface.}
    \label{fig:NPW_cases}
\end{figure}
A first challenge in this modeling effort concerns the \textit{accuracy of the refractive index itself}. 
Accurate EM propagation prediction relies on reliably reconstructing the atmospheric refractive index from weather variables such as pressure, temperature, relative humidity, hydrometeors, and rainfall~\cite{liebe1983atmospheric, liebe1993propagation, series2019attenuation, series2019refractivity}. 
However, conventional reconstruction approaches---(i) measurement-driven interpolation and (ii) remote-sensing estimation---often struggle to provide physically consistent, high-resolution latitude--longitude refractivity profiles required for LEO link analysis~\cite{de2020radiometric, luini2017attenuation, jang2024kriging, wang2021prediction, battaglia2019gpm, kou2023simulation, zhang2023estimation}.\footnote{The former is limited by the sparseness of radiosonde stations, while the latter frequently lacks vertically resolved data.}
A second challenge arises from fundamental physical issue. 
In lossy atmospheric media, an uniform plane wave no longer preserves its original structure. 
The complex-valued refractive index caused by hydrometeors and absorption forces the wave to evolve into a \textit{non-uniform plane wave}---i.e., the directions of phase progression and attenuation deviate from each other, as illustrated in Fig.~\ref{fig:NPW_cases}. 
Such non-uniformity can modify both the refraction angle and the accumulated path loss \cite{kim2023critical}. 
However, existing studies have not explicitly accounted for the non-uniform behavior of plane waves in inhomogeneous and lossy atmospheric media.
Instead, most previous EM propagation studies implicitly assume uniform plane waves---without a clear physical justification---which may overlook the intrinsic non-uniform propagation mechanisms that arise in lossy refractivity profiles.

With these two challenges in mind, it is natural to ask: \textit{Would non-uniform plane waves meaningfully change boresight error or path-loss estimates compared with the uniform plane-wave approximation for LEO links extending hundreds of kilometers?}
This work addresses this question by developing an analytical and computational framework capable of quantifying the influence of plane wave non-uniformity in inhomogeneous complex-index atmospheres. 
First, we incorporate numerical weather prediction (NWP) data into the MPM93 model---including cloud, precipitation, and land–sea distributions~\cite{kma_apihub}---to reconstruct a physically well-grounded, high-resolution, complex-valued 3D atmospheric refractivity map.
Second, we propose a fast and numerically stable ray-tracing framework suitable for inhomogeneous complex-index media with extremely small loss.
This resolves the numerical instability inherent in the complex-Snell non-uniform plane-wave formulation of Chang \textit{et al.}~\cite{chang2005ray}, which---despite its wide adoption~\cite{foley2014inhomogeneous, ballington2024light, lameirinhas2024plasmonic}---suffers from severe subtractive cancellation in attenuation-related terms under standard double precision, rendering atmospheric simulations unreliable.
Third, using the developed computational framework, we quantitatively assess the validity of the conventional uniform plane-wave model, providing simulation-based guidance for practical LEO communication scenarios.

Throughout this work, we adopt the time convention $e^{j\omega t}$.

\section{NWP-based atmospheric refractivity modeling}
\begin{figure}[t]
    \centering
    \begin{subfigure}[b]{0.48\linewidth}
        \includegraphics[width=\linewidth]{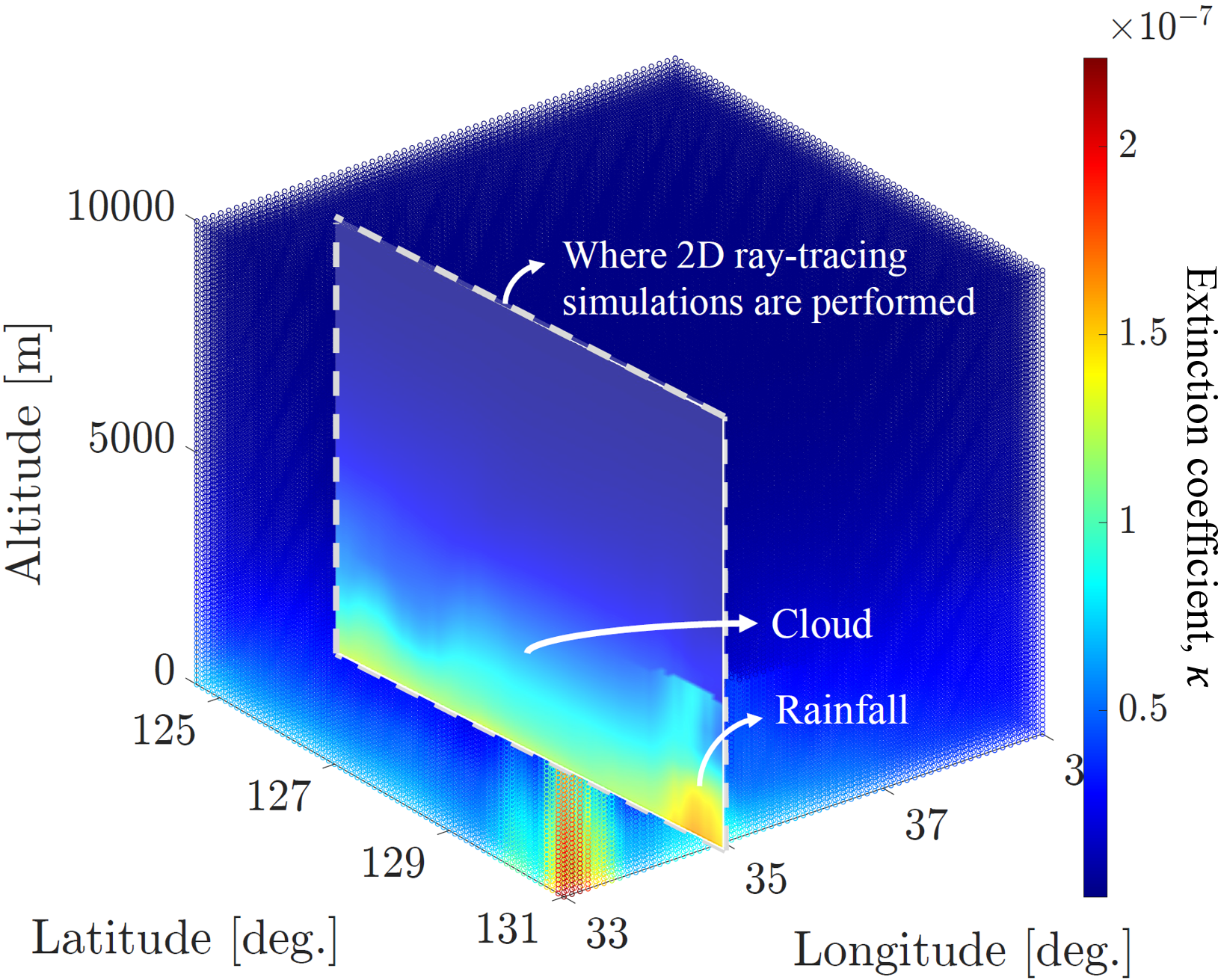}
        \caption{3D $\kappa$ map from KIM}
        \label{fig:KIM_RH}
    \end{subfigure}
    \hfill
    \begin{subfigure}[b]{0.47\linewidth}
        \includegraphics[width=\linewidth]{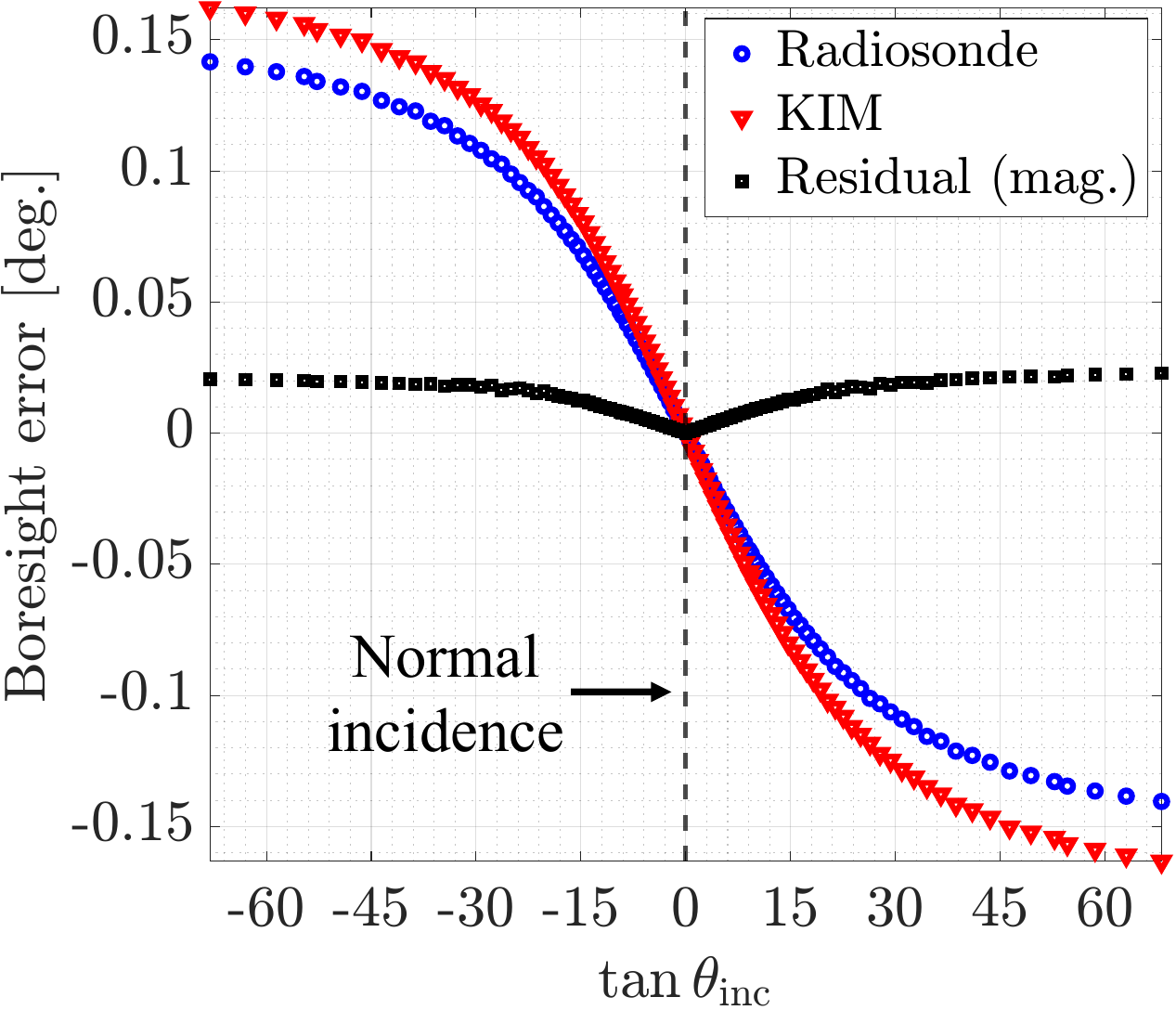}
        \caption{Boresight error}
        \label{fig:KIM_W}
    \end{subfigure}
    \caption{(a) 3D $\kappa$ map from the KIM data and (b) boresight error for the two atmospheric refractivity models.}    
\end{figure}

The atmosphere over South Korea is modeled as a stratified medium up to 10~km, encompassing the troposphere where most weather-driven variations occur. 
Within this region, refractivity changes significantly with altitude and meteorological conditions, leading to pronounced refraction effects.
Considering an 18-GHz link, the complex refractivity $\tilde{N}$\footnote{The relationship between $\tilde{n}$ and $\tilde{N}$ is given by $\tilde{n} = \tilde{N} \times 10^{-6} + 1 = n - j\kappa$ \cite{kim2021IEEEAccess}, where $n$ and $\kappa$ denote the refractive index and the extinction coefficient, respectively.} is retrieved via MPM93 as the sum of dry air, water vapor, cloud, and rain components \cite{liebe1983atmospheric,liebe1993propagation}:
$\tilde{N} = \tilde{N}_{\mathrm{dry\,air}} + \tilde{N}_{\mathrm{water\,vapor}} + \tilde{N}_{\mathrm{cloud/fog}} + \tilde{N}_{\mathrm{rainfall}}.$
\begin{figure}[t] 
  \centering
  \begin{subfigure}[b]{0.45\linewidth}
    \includegraphics[width=\linewidth, trim={0 1.6cm 0 2.1cm}, clip]{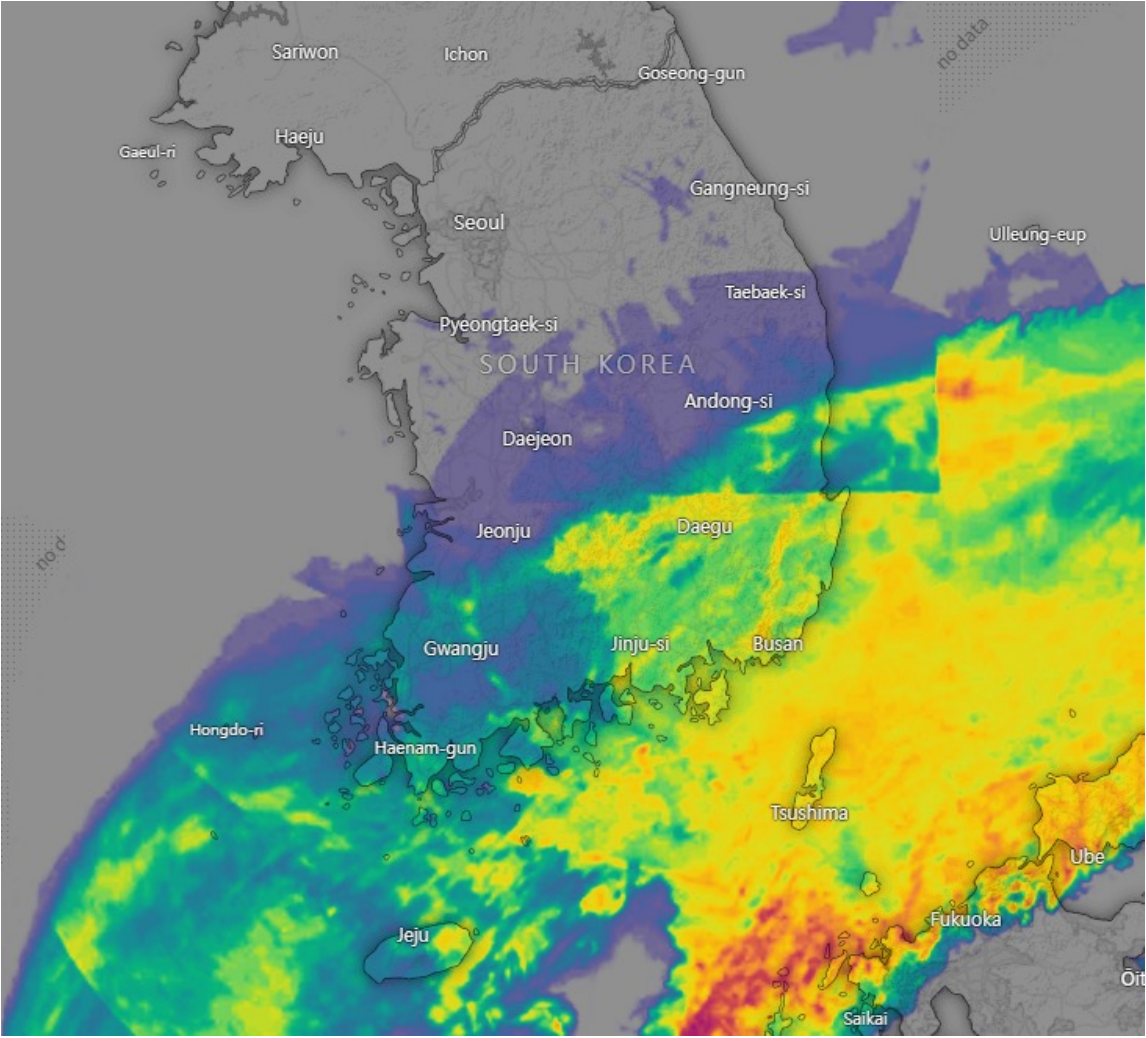}
    \caption{Precipitation}
    \label{fig:windy}
  \end{subfigure}
  \hfill
  \begin{subfigure}[b]{0.48\linewidth}
    \includegraphics[width=\linewidth, trim={0 1.6cm 0 2.1cm}, clip]{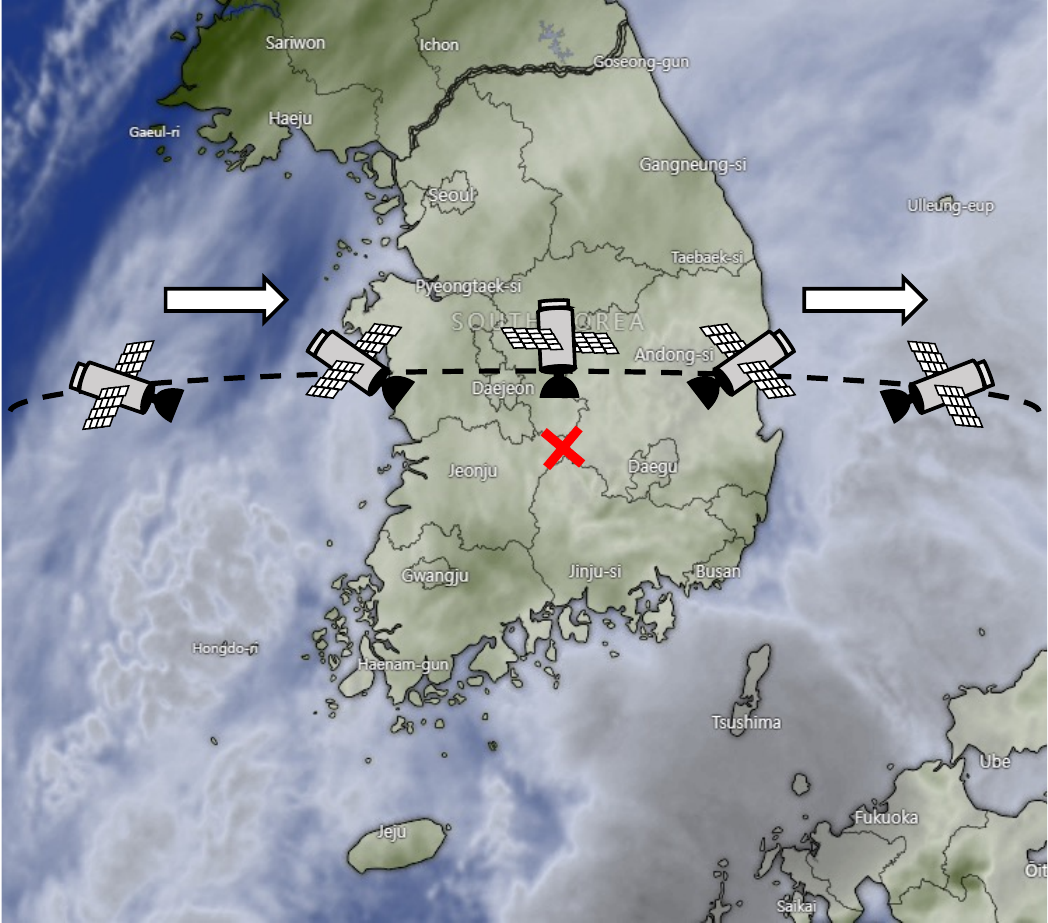}
    \caption{Clouds distribution}
    \label{fig:radiosonde_windy}
  \end{subfigure}
\caption{Windy weather forecast~\cite{windy} on 2 November 2024, 00:00 KST: (a) precipitation and (b) cloud distribution.}
\label{fig:windy}
\end{figure}
\begin{figure}[t] 
  \centering
  \begin{subfigure}[b]{0.49\linewidth}
    \includegraphics[width=\linewidth]{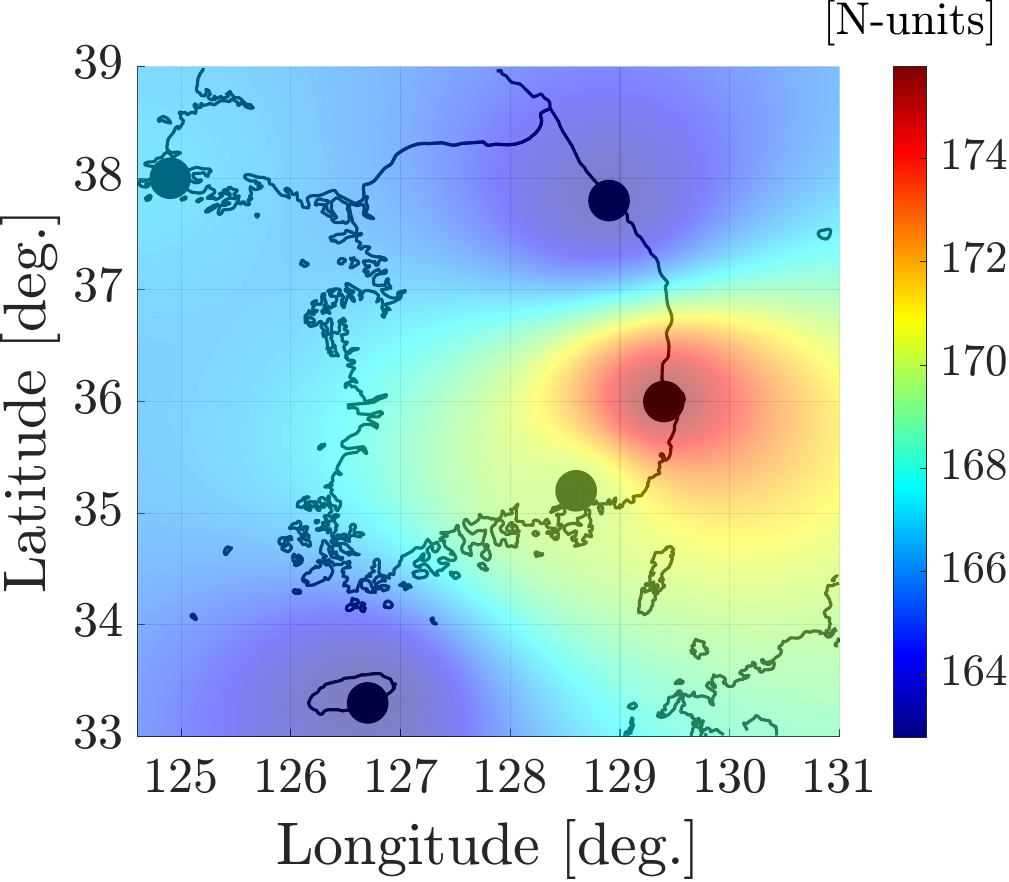}
    \caption{Radiosonde}
    \label{fig:radiosonde}
  \end{subfigure}
  \hfill
  \begin{subfigure}[b]{0.49\linewidth}
    \includegraphics[width=\linewidth]{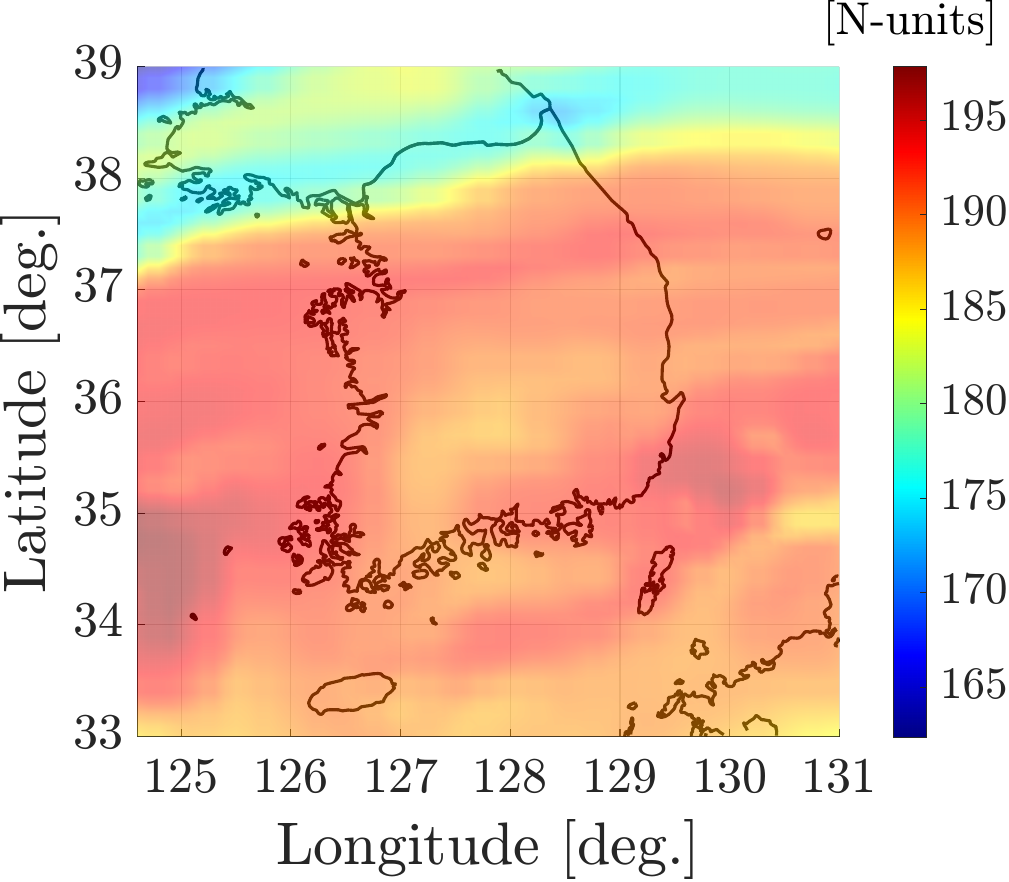}
    \caption{KIM}
    \label{fig:KIM}
  \end{subfigure}
  \caption{Real part of complex atmospheric refractivity at 5~km altitude over South Korea, reconstructed from (a)  radiosonde stations (black circles) using IDW and (b) KIM data.}
  \label{fig:KIM_inputs}
\end{figure}
For analysis, we selected 2 November 2024 (00:00 KST), during Typhoon Kong-rey, to validate the model under highly inhomogeneous atmospheric conditions. 
Fig.~\ref{fig:windy} shows the corresponding weather-forecast precipitation and cloud distributions obtained from the Windy platform~\cite{windy}.
To reconstruct the atmospheric refractivity, we first apply inverse-distance-weighting (IDW) interpolation in latitude using radiosonde measurements from stations 47102, 47104, 47138, 47155, and 47186, as shown in Fig.~\ref{fig:radiosonde}.
However, because the station network is sparse, this simple interpolation scheme is fundamentally limited and produces refractivity fields that deviate noticeably from those inferred in Fig.~\ref{fig:windy}.
To address this limit, we instead use the Korea Integrated Model (KIM), a NWP model optimized for the Korean Peninsula, to obtain a high-resolution refractivity field.\footnote{Depending on the domain, the KIM model provides horizontal resolutions of $0.1^\circ$  in both latitude and longitude over Korea, corresponding to grid spacings of approximately 11~km.}
As shown in Fig.~\ref{fig:KIM}, the resulting refractivity field exhibits significantly higher consistency with the Windy data.\footnote{The KIM dataset provides geopotential height $HGT$~[m], relative humidity $RH$~[\%], temperature $T$~[K], cloud liquid/ice mixing ratios ($TQC$, $TQI$)~[kg/kg] on 31 isobaric levels, and accumulated precipitation $PREC$~[kg/m$^{2}$]. Altitude-dependent pressure $P$~[Pa] is obtained by first-order interpolation of the isobaric fields. The accumulated precipitation is converted to an hourly rainfall rate $R$~[mm/h] using a $\pm 3$~h window, and rainfall-induced refractivity is evaluated using a uniform precipitation layer~\cite{series2013rain}. Cloud hydrometeors are represented by $W$~[g/m$^{3}$], computed from $TQC$, $TQI$, and the dry-air density $\rho_{\mathrm{dry}} = 100P/(R_{\mathrm{dry}}T)$, giving $W_{\mathrm{liquid}}=\rho_{\mathrm{dry}}TQC$ and $W_{\mathrm{ice}}=\rho_{\mathrm{dry}}TQI$. Finally, isobaric data were interpolated onto altitude coordinates to form a three-dimensional stratified refractivity field over latitude, longitude, and altitude.}
The extinction coefficient $\kappa$ in the KIM-based refractivity model ranges from $4.79\times10^{-10}$ to $2.24\times10^{-7}$, spanning light to heavy rainfall conditions.
Even under strong precipitation, $\kappa$ remains several orders of magnitude smaller than $n$, so the atmosphere is only weakly lossy; nevertheless, the presence of loss still produces a sufficiently complex medium capable of introducing plane-wave non-uniformity.
To assess how different refractivity models influence ray behavior, we perform downlink ray-tracing simulations in which the receiver is fixed at the central longitude of our analysis area ($35.6^{\circ},\mathrm{N},\ 127.8^{\circ},\mathrm{E}$).
Rays are launched from a satellite located at various longitudes at an altitude of 10 km with incident angles
$\theta_{\mathrm{inc}} \in [-90^{\circ}, 90^{\circ}]$, defined by the elevation angle between the transmitter and receiver, while all rays are directed toward the receiver, as illustrated in Fig.~\ref{fig:radiosonde_windy}.
Fig.~\ref{fig:KIM_W} presents the resulting boresight-angle error as a function of incident angle when the uniform-plane-wave approximate ray tracer is applied to refractivity profiles reconstructed from both radiosonde and KIM data.\footnote{The approximation assumes that the phase and attenuation directions are aligned, with refraction and attenuation governed solely by the real and imaginary parts of the refractive index, respectively.}
It can be observed that the deviation between the two models increases as the rays approach more grazing incidence.
In particular, the KIM-based model exhibits larger boresight-angle errors at relatively higher (i.e., less grazing) incident angles, whereas the radiosonde-based model produces noticeably smaller errors at the same incident angles.
The boresight-error discrepancies arise from differences in refractivity fidelity: the sparse radiosonde network produces coarse horizontal gradients, whereas the NWP-based KIM model captures much finer spatial structure.
These results underscore the sensitivity of refraction predictions to refractivity accuracy and highlight the need for high-resolution, physically consistent atmospheric models.

\section{Proposed non-uniform Ray-Tracing Algorithm}
When a plane wave encounters an interface between two media with complex-valued refractive indices
($\tilde{n}_1 = n_1 - j\kappa_1$ and $\tilde{n}_2 = n_2 - j\kappa_2$), the directions of phase propagation and amplitude attenuation no longer coincide~\cite{balanis2012advanced}.
However, applying the non-uniform plane-wave ray-tracing formulation of~\cite{chang2005ray} to atmospheric conditions---where the loss terms are extremely small---reveals that standard double-precision arithmetic suffers from severe subtractive cancellation, destabilizing the evaluation of attenuation-related quantities and preventing reliable simulation.
While long-double precision alleviates this issue, it incurs substantial computational overhead.
To achieve both accuracy and efficiency, we develop a new fast and numerically stable non-uniform plane-wave ray-tracing algorithm that remains robust even in standard double precision.
Following~\cite{chang2005ray}, the apparent refractive indices $N_m$ and $K_m$ are introduced to express the complex wave vector in medium $m$ as 
$\tilde{\mathbf{k}}_{p} = k_{0}(N_{m}\mathbf{e}_{p} - jK_{m}\mathbf{f}_{p})$ 
for the incident ($p=\mathrm{i}$), reflected ($p=\mathrm{r}$), and transmitted ($p=\mathrm{t}$) waves, where $\mathbf{e}_{p}$ and $\mathbf{f}_{p}$ denote the phase- and attenuation-direction unit vectors, as defined in Fig. \ref{fig:IHPW incidence}.\footnote{In lossy media, the phase and attenuation directions differ, so the phase-matching condition is enforced by redefining the effective components via the apparent refractive indices. Consequently, an ordinary wave vector $\mathbf{k}=k_0(n-j\kappa)\mathbf{e}_p$ is replaced by the modified one $\tilde{\mathbf{k}}_p$ while preserving $\lvert \mathbf{k} \rvert = \lvert \tilde{\mathbf{k}}_p \rvert$.
}

The key idea of our algorithm is to reformulate the non-uniform plane-wave interface relations into numerically stable expressions so that all quantities involving extremely small loss terms can be evaluated reliably in standard double precision. 
We stabilize the computation of $(N_1, K_1)$ in medium~I and solve the medium~II refraction problem using reduced quartic forms whose physically admissible roots remain numerically well conditioned even when $\kappa \ll n$.
We consider 2D incidence, corresponding to the plane-of-incidence polarization in 3D. In medium~I, the NPW dispersion relation and boundary conditions determine $N_1$ and $K_1$. Given $(\tilde{n}_1, \tilde{n}_2)$ and incident angles $(\theta_{\mathrm{i}}, \psi_{\mathrm{i}})$, Chang’s formulation evaluates $K_m = \sqrt{N_m^2 - n_m^2 + \kappa_m^2}$, but because $\kappa \ll n$, the relation $N \approx n$ leads to severe subtractive cancellation in the computation of $K_m$.
To avoid cancellation, we define $a = n_1^2 - \kappa_1^2$, $b = n_1\kappa_1$, $c = \cos(\theta_{\mathrm{i}} - \psi_{\mathrm{i}})$, and $r = ac/b$, and compute $K_1^2$ using stable expressions. When $c$ is well behaved, $K_1^2 = 0.5\, a\, [2b/(ac)]^2 \bigl(\sqrt{1 + (2b/ac)^2} + 1\bigr)^{-1}$; when $b$ is well behaved, $K_1^2 = 2a \bigl(r^2 + r\sqrt{r^2 + 4}\bigr)^{-1}$; and in extreme limits, $K_1^2 \approx a/r^2$ for $r \gg 1$ and $K_1^2 \approx a/r$ for $r \ll 1$. The normal component then follows as $N_1^2 = K_1^2 + a$.
In medium~II, the NPW relations reduce to the quartic forms 
$N_2^4 - (A + B)\,N_2^2 + AB - D^2 = 0$ and 
$K_2^4 - (A - B)\,K_2^2 - AB - D^2 = 0$, 
where 
$A = N_s^2 + K_s^2$, 
$B = n_2^2 - \kappa_2^2$, 
$D = n_2\kappa_2 - N_s K_s$, 
$N_s = N_1\sin\theta_{\mathrm{i}}$, 
$K_s = K_1\sin\psi_{\mathrm{i}}$, 
and the physically valid branch is selected by enforcing $K_2 = \sqrt{N_2^2 - n_2^2 + \kappa_2^2}$.
Once $(\theta_{\mathrm{t}}, \psi_{\mathrm{t}})$ are obtained, the transmitted field is updated using NPW Fresnel coefficients, and the field entering the next layer is 
$\mathbf{E}_{i}^{q+1} = \mathbf{E}_t^{q}\,
\exp[-k_0 K_2 \cos(\theta_{\mathrm{t}} - \psi_{\mathrm{t}})\,l_2]\,
\exp[-j k_0 N_2 l_2]$,
where $l_2$ is the propagation distance and $q$ is the iteration index.

\begin{figure}[t]
  \centering
  \begin{subfigure}[b]{0.47\columnwidth} 
    \includegraphics[width=\linewidth]{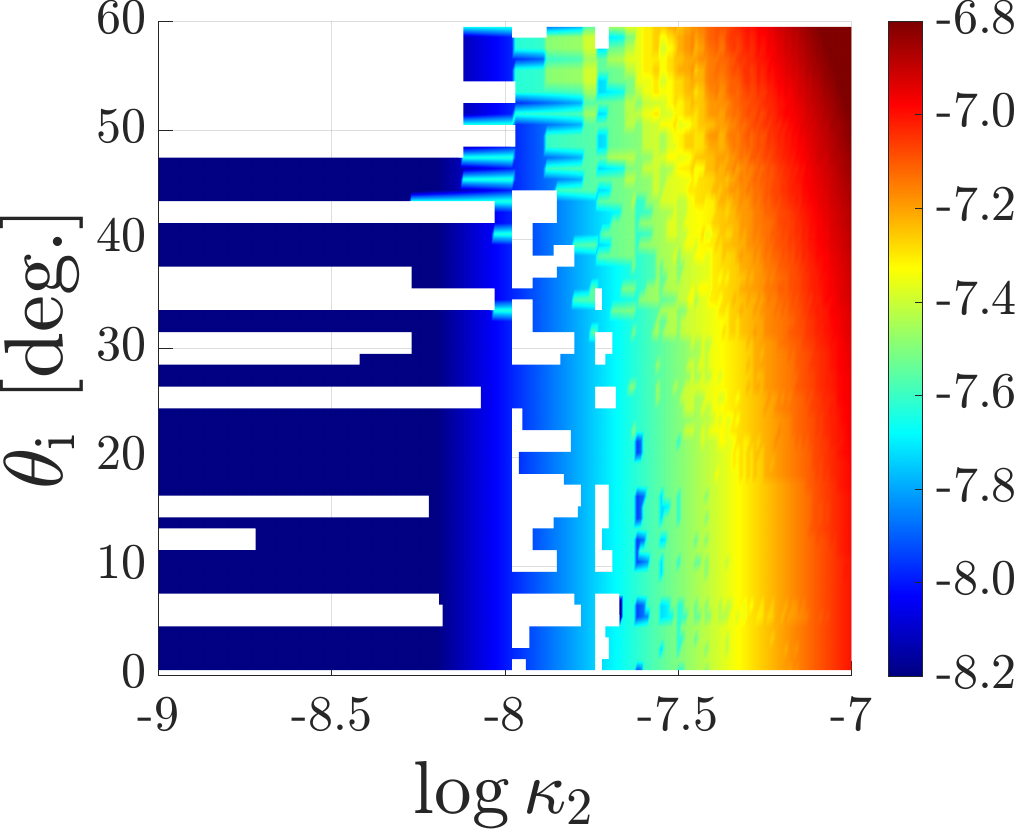}
    \caption{$\log K_{2}$}
    \label{fig:chang_K2}
  \end{subfigure}
  \hfill
  \begin{subfigure}[b]{0.47\columnwidth} 
    \includegraphics[width=\linewidth]{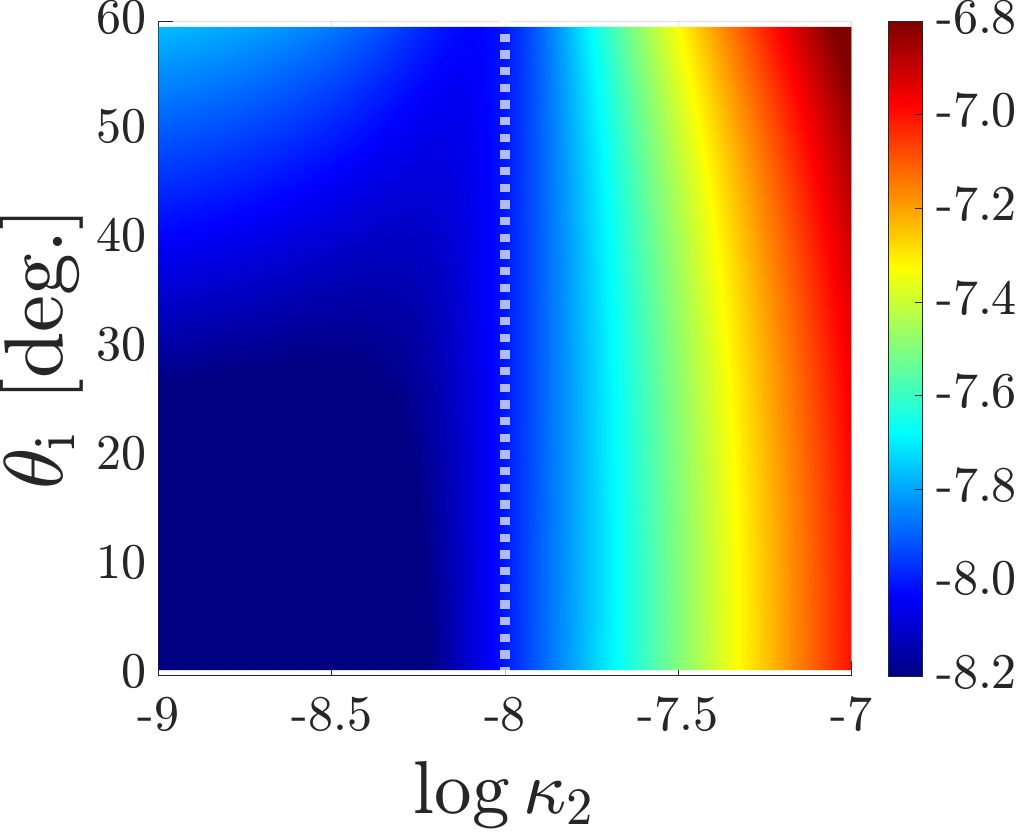}
    \caption{$\log K_{2}$}
    \label{fig:ours_delta}
  \end{subfigure}
  \begin{subfigure}[b]{0.47\columnwidth} 
    \includegraphics[width=\linewidth]
    {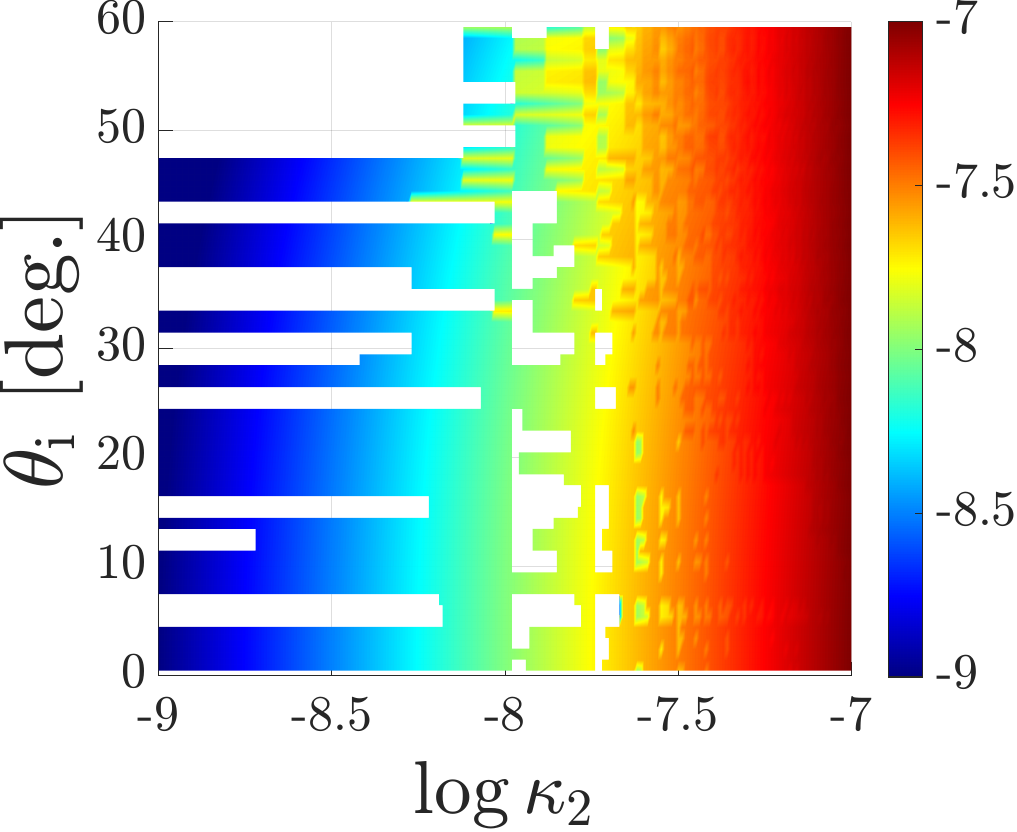}
    \caption{$\log ( K_{2}\cos \alpha_{\mathrm{t}} )$}
    \label{fig:chang_delta}
  \end{subfigure}
  \hfill
  \begin{subfigure}[b]{0.47\columnwidth} 
    \includegraphics[width=\linewidth]
    {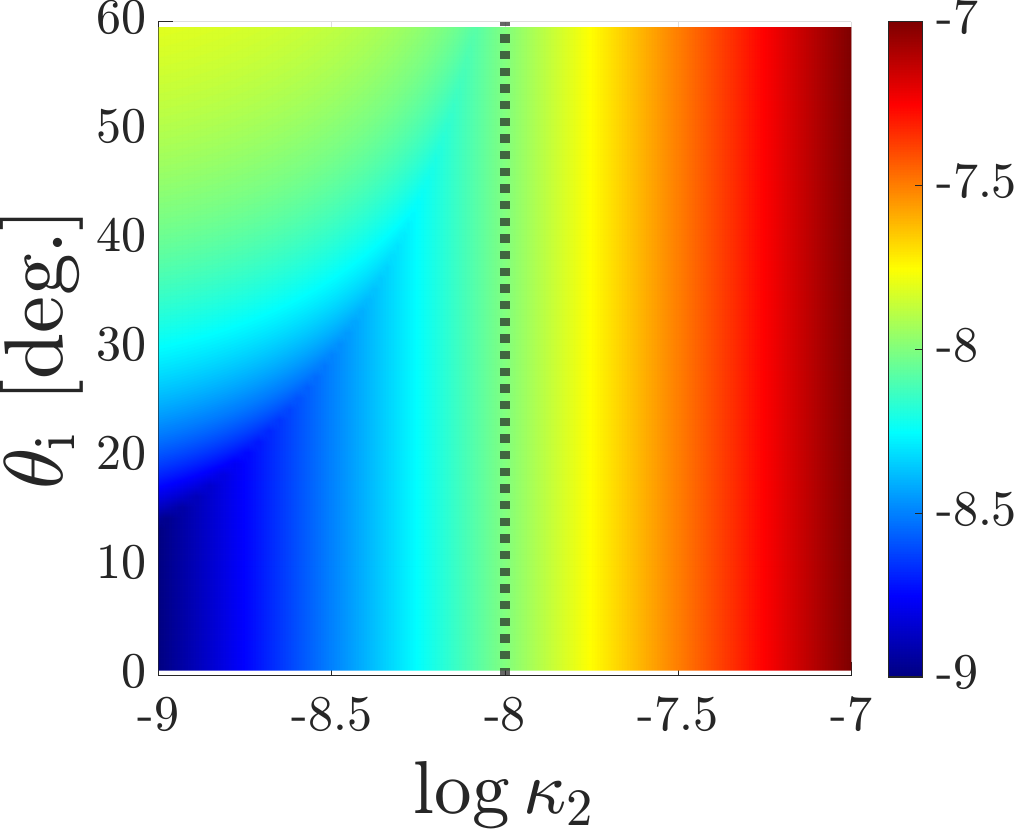}
    \caption{$\log ( K_{2}\cos\alpha_{\mathrm{t}} )$}
    \label{fig:ours_K2cos}
  \end{subfigure}
  \caption{Results using Chang’s method: (a), (c); and using the proposed method: (b), (d). NPW with $\alpha_{\mathrm{i}} = 10^{\circ}$ is incident and refracts into the second medium. Vertical dashed lines in (b) and (d) denote the fixed value $\kappa_{1}=10^{-8}$.}
  \label{fig:comparison_3x2}
\end{figure}
\section{Numerical Results}

\subsection{Single interface refraction}
\begin{figure}[t]
  \centering
  \begin{subfigure}[b]{0.47\columnwidth} 
    \includegraphics[width=\linewidth]{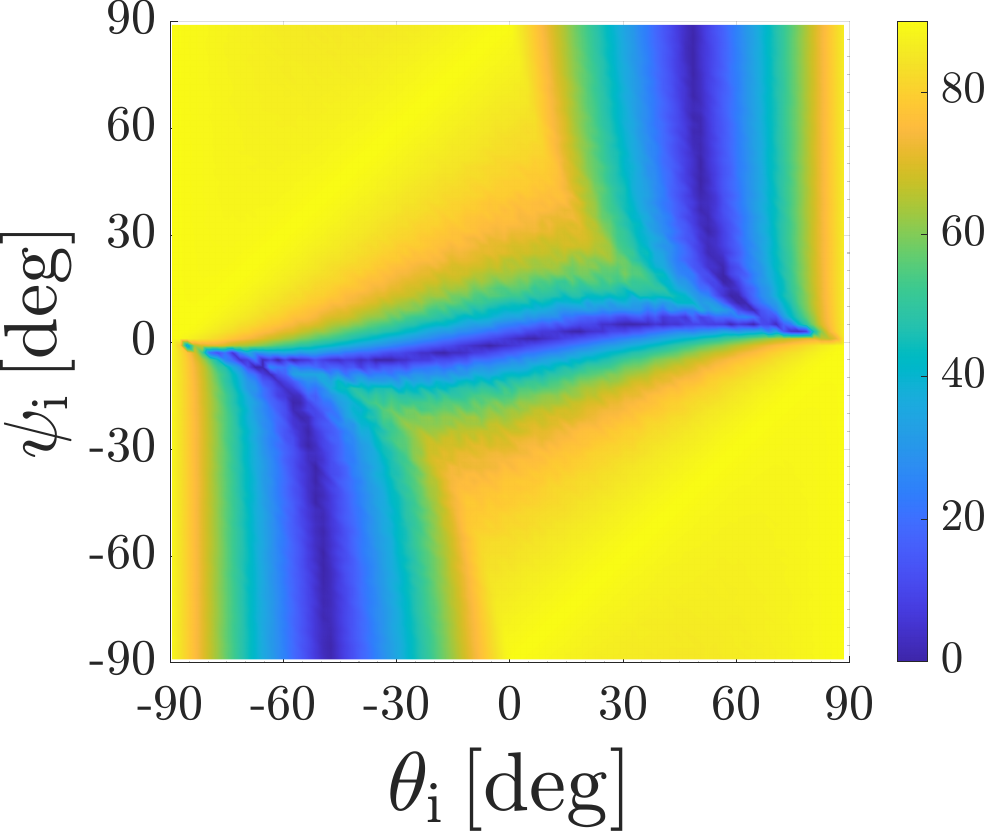}
    \caption{$\alpha_{\mathrm{t}}$}
    \label{fig:chang_K2}
  \end{subfigure}
  \hfill
  \begin{subfigure}[b]{0.47\columnwidth} 
    \includegraphics[width=\linewidth]{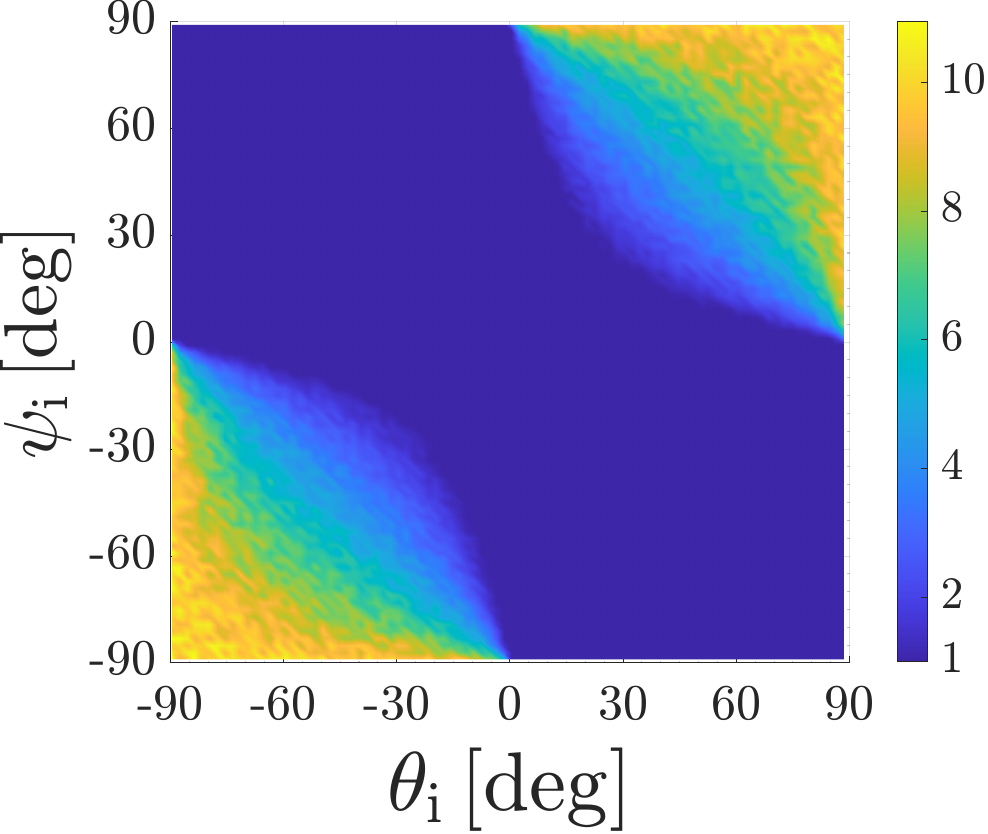}
\caption{$K_{2}\cos{\left(\alpha_{\mathrm{t}}\right)}/\kappa_{2}$}
    \label{fig:chang_delta}
  \end{subfigure}
  \begin{subfigure}[b]{0.47\columnwidth} 
    \includegraphics[width=\linewidth]
    {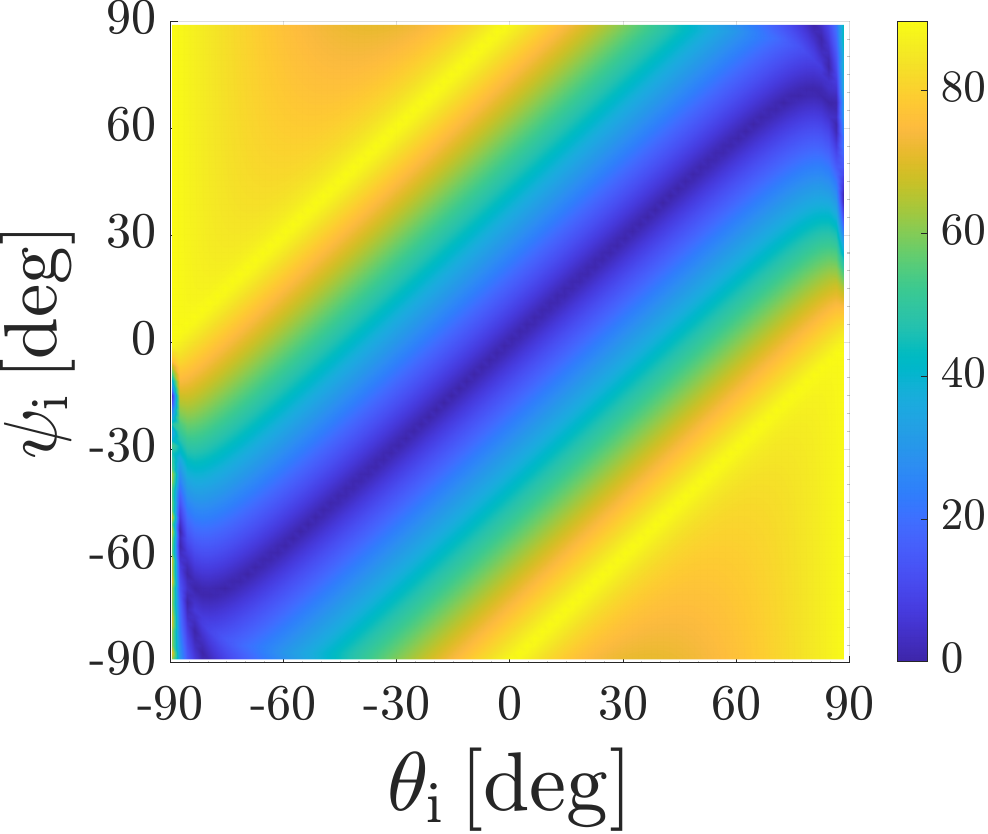}
    \caption{$\alpha_{\mathrm{t}}$}
    \label{fig:ours_delta}
  \end{subfigure}
  \hfill
  \begin{subfigure}[b]{0.47\columnwidth} 
    \includegraphics[width=\linewidth]
    {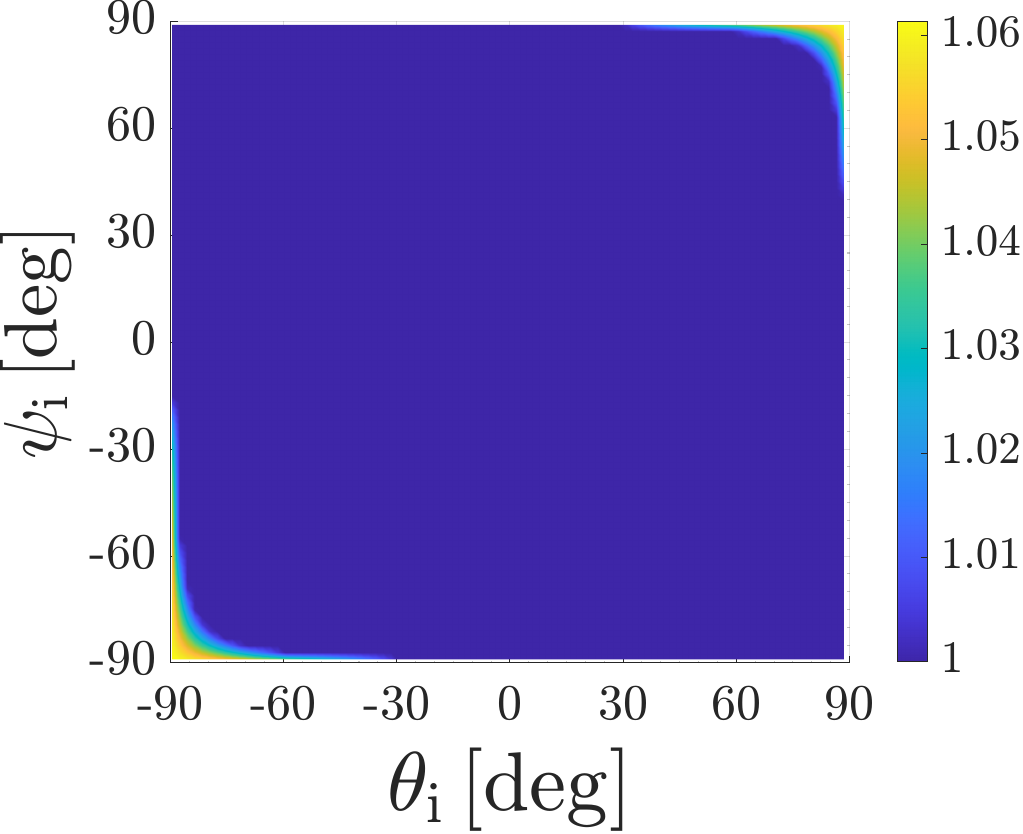}
\caption{$K_{2}\cos{\left(\alpha_{\mathrm{t}}\right)}/\kappa_{2}$}
    \label{fig:ours_K2cos}
  \end{subfigure}
\caption{$\alpha_t$ in (a), (c) and $K_2 \cos\alpha_t / \kappa_2$ in (b), (d) for both $n_1/n_2 \approx 1$. Parameters: (a)--(b) $\kappa_1 = 1.94\times10^{-10}$, $\kappa_2 = 3.58\times10^{-11}$; (c)--(d) $\kappa_1 = 1.54\times10^{-8}$, $\kappa_2 = 1.32\times10^{-8}$. The ratio approaching unity confirms $K_2 \cos\alpha_t \approx \kappa_2$.}
\label{fig:validation}
\end{figure}

To examine the non-uniformity of plane waves under atmospheric-scale complex refractivity, we analyze the wave refraction at a single interface across a range of incident angles. 
We set $n_1 = 1.0001$, $n_2 = 1.0002$, and $\kappa_{1}=10^{-8}$, while varying $\kappa_{2}$ from $10^{-7}$ to $10^{-9}$ for different incident angles, and investigate the resulting refraction characteristics through $K_{2}$ and the effective attenuation term $K_{2}\cos\alpha_{t}$.
Fig.~\ref{fig:comparison_3x2} compares Chang’s method with the proposed ray tracer, with both implementations evaluated in standard double precision.
Chang’s method suffers from severe subtractive cancellation, yielding incorrect attenuation estimates, whereas the proposed ray-tracing maintains numerical stability even for extremely small $\kappa_{2}$.
A notable observation is that when $\kappa_{2} \ge \kappa_{1}$, the relation $\kappa_{2} \approx K_{2}\cos\alpha_{t}$ continues to hold even though $\alpha_{t} \neq 0$. 
In this regime, the increase in $\alpha_{t}$ is compensated by a corresponding change in $K_{2}$, preserving the effective attenuation and reproducing the phase-path loss predicted by the uniform-plane-wave approximation. 
Conversely, when $\kappa_{2} < \kappa_{1}$, such compensation no longer occurs, i.e., $\kappa_{2} \neq K_{2}\cos\alpha_{t}$, and the attenuation predicted by the uniform approximation deviates from the result by the proposed algorithm. 
Randomized tests enforcing spatial continuity further confirm that when the loss terms are comparable (e.g., ratio $\approx 0.97$), the uniform approximation remains accurate, as illustrated in Fig.~\ref{fig:validation}.

\subsection{LEO downlink ray-tracing simulations}
\begin{figure}
    \centering
    \includegraphics[width=1\linewidth]{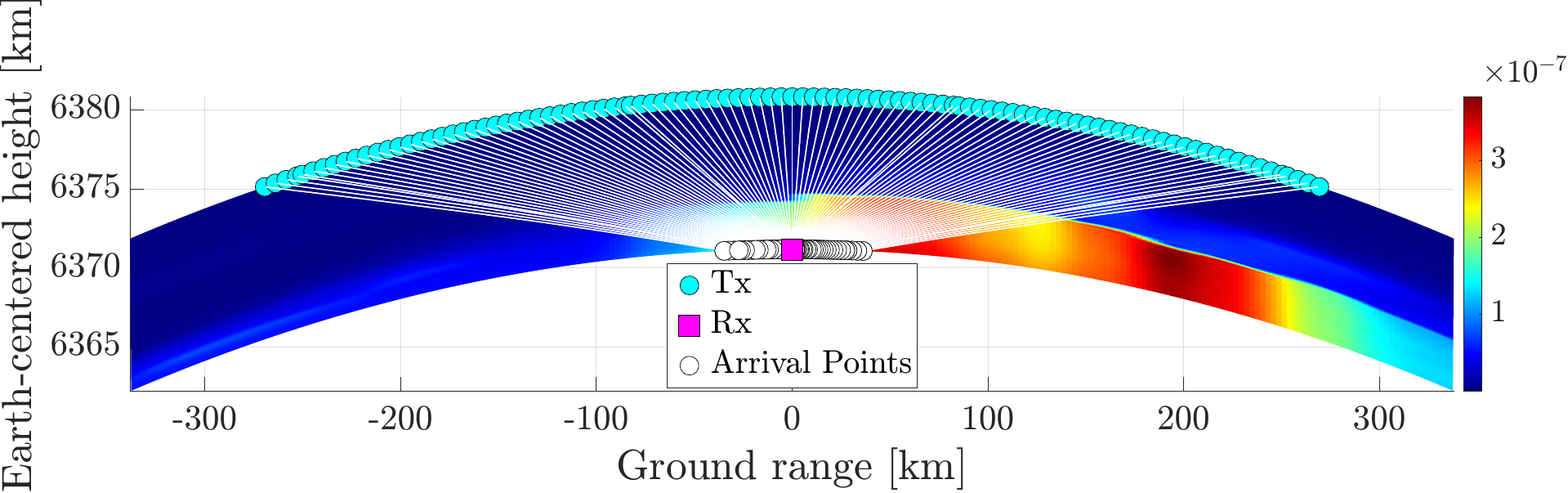}
    \caption{Ray paths upon the extinction coefficient $\kappa$ map on 21 September 2024, 12:00 KST, under heavy rainfall.}
    \label{fig:analysis_domain}
\end{figure}

\begin{figure}[t] 
  \centering
  \begin{subfigure}[b]{0.45\linewidth}
    \includegraphics[width=\linewidth]{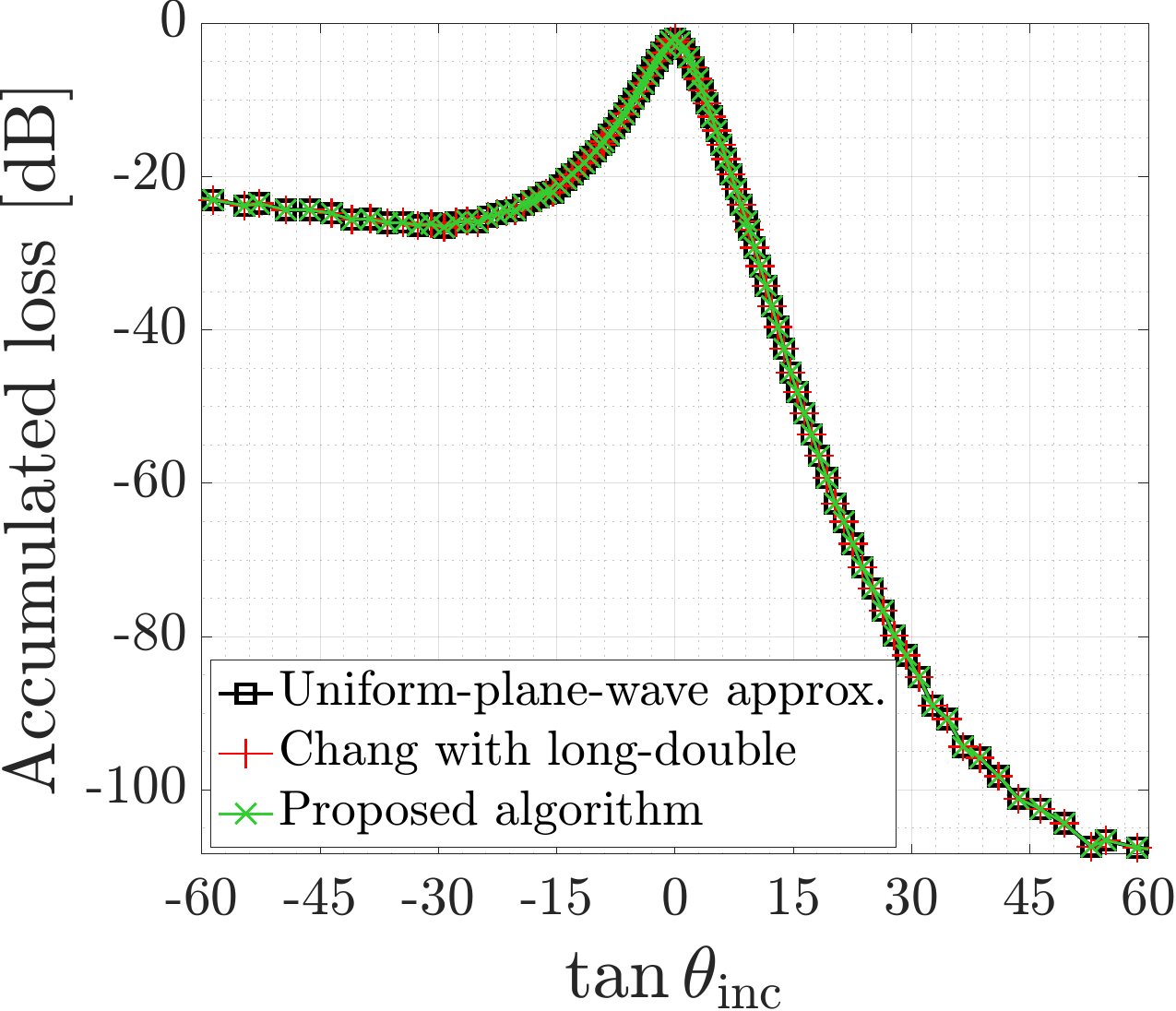}
    \caption{Accumulated loss}
    \label{fig:loss}
  \end{subfigure}
  \hfill
  \begin{subfigure}[b]{0.47\linewidth}
    \includegraphics[width=\linewidth]{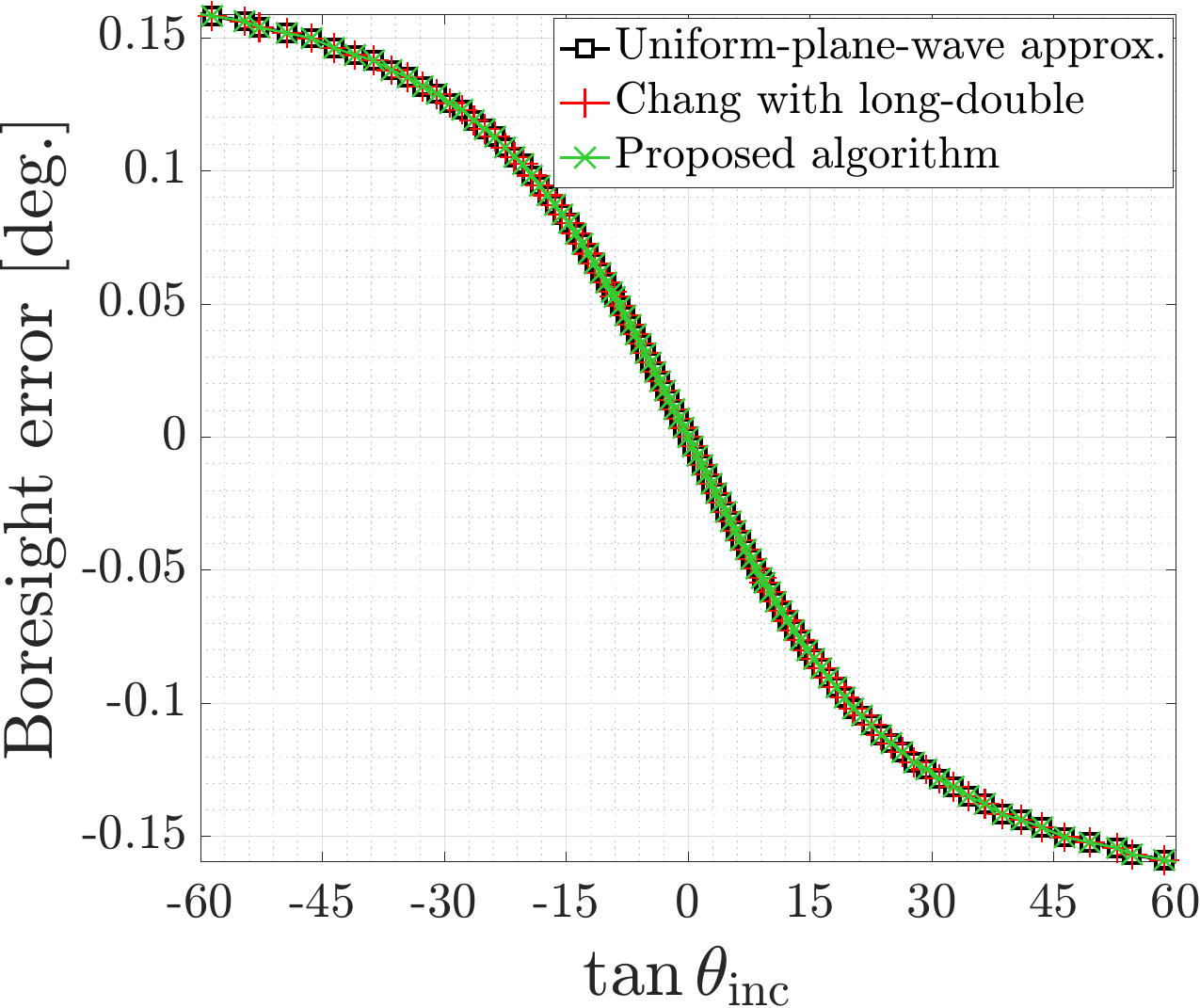}
    \caption{Boresight error}
    \label{fig:boresight}
  \end{subfigure}
    \caption{Simulation results from three different methods.}
\label{fig:LEO_sim_result}
\end{figure}

We evaluate LEO downlink performance by applying three methods to 2D cross 
sections extracted from the 3D atmospheric refractivity field:  
(i) the uniform-plane-wave approximation, (ii) Chang's method with long-double precision, and 
(iii) the proposed algorithm, as illustrated in Fig.~\ref{fig:analysis_domain}.  
As shown in Fig.~\ref{fig:LEO_sim_result}, all three methods produce nearly 
identical refraction paths and attenuation profiles.
Because $\kappa \ll n$, refraction is effectively determined by the real part of 
the refractive index, and the slowly varying complex refractivity under weak 
loss is well captured by the uniform approximation.  
Our results demonstrate that the non-uniformity of plane waves has negligible influence 
on LEO propagation: the refraction angle is insensitive to the loss term, and 
the accumulated attenuation remains unchanged.  
This occurs because the effective loss factor $K\cos\alpha$ closely matches the 
intrinsic loss $\kappa$, as variations in $\cos\alpha$ compensate differences 
between $K$ and $\kappa$.  
Thus, even under heavy rainfall, the uniform plane-wave approximation provides accurate 
boresight-error and path-loss estimates while enabling fast computation.  
In contrast, using long-double precision in Chang’s formulation preserves 
accuracy but incurs significant computational cost.  
The proposed algorithm, by contrast, achieves the same accuracy using standard double precision.  
Both the uniform approximation and the proposed algorithm scale as $O(1)$ per evaluation. 
Although the proposed method incurs a small overhead due to the iterative selection of the physically admissible root of the quadratic equation for $K_2$.
Across 1{,}000 repeated evaluations, the uniform-plane-wave model required an average of 0.1948~ms per evaluation (standard deviation 0.0271~ms), whereas the proposed algorithm required 0.2065~ms 
(standard deviation 0.0274~ms), corresponding to only about a 6\% slowdown.  
In contrast, Chang’s long-double implementation took 4.8888~ms on 
average with a standard deviation of 0.5468~ms, making it roughly 24 times slower than the proposed method. 
Thus, the new algorithm achieves numerical stability comparable to long-double precision while maintaining a runtime essentially indistinguishable from the uniform-plane-wave approximation.

\section{Conclusion}
We have developed a high-resolution atmospheric refractivity model reconstructed from numerical weather prediction (NWP) data, together with a fast and numerically stable ray-tracing algorithm that explicitly accounts for the non-uniformity of plane waves. 
Our analysis shows that, under most practical operating conditions, the non-uniformity effect is sufficiently small that it may be safely neglected without compromising the accuracy of boresight-error or path-loss estimates. 
Future work will focus on improving spatial resolution beyond the native limit $0.1^\circ$ of NWP data by employing physics-informed super-resolution techniques based on transformers and physics-informed neural networks~\cite{li2024deepphysinet}.


\newpage
\bibliographystyle{IEEEtran}
\bibliography{reference}

\end{document}